\documentclass[fleqn,11pt]{bcl_manuscript}
\usepackage{soul}
\usepackage{setspace}
\usepackage[capitalize]{cleveref}
\title{Constructor algorithms for building unconventional computers able to solve NP-complete problems}

\author[1,*]{Tony McCaffrey}
\author[2]{Thomas E. Gorochowski}
\author[3,4]{Lee Spector}

\affil[1]{~Department of Computer Science, Eagle Hill School, Massachusetts, USA}
\affil[2]{~School of Biological Sciences, University of Bristol, Bristol, UK}
\affil[3]{~Department of Computer Science, Amherst College, Massachusetts, USA}
\affil[4]{~College of Information and Computer Sciences, University of Massachusetts, Amherst, Massachusetts, USA}

\corrauthor[]{T.M. (tony@constructoralgorithms.com)}

\begin{document}

\setlength{\abovedisplayskip}{3pt}
\setlength{\belowdisplayskip}{3pt}
\flushbottom
\maketitle
\thispagestyle{empty}

\section*{Abstract}

Nature often builds physical structures tailored for specific information processing tasks with computations encoded using diverse phenomena. These can sometimes outperform typical general-purpose computers. However, describing the construction and function of these unconventional computers is often challenging. Here, we address this by introducing constructor algorithms in the context of a robotic wire machine that can be programmed to build networks of connected wires in response to a problem and then act upon these to efficiently carry out a desired computation. We show how this approach can be used to solve the NP-complete Subset Sum Problem (SSP) and provide information about the number of solutions through changes in the voltages and currents measured across these networks. This work provides a foundation for building unconventional computers that encode information purely in the lengths and connections of electrically conductive wires. It also demonstrates the power of computing paradigms beyond digital logic and opens avenues to more fully harness the inherent computational capabilities of diverse physical, chemical and biological substrates.

\section*{Introduction}

General-purpose electronic computers built on the foundation of digital logic are ubiquitous in everyday life. Their highly-programmable nature makes them ideal for a wide variety of tasks. However, this flexibility often comes at a cost --- their efficiency in performing specific types of computation \cite{Hameed2010}. In contrast, biological systems display a far wider diversity in the ways that information is encoded and processed \cite{Bourret2002, Brody2023, Berridge2001, Adamatzky2018, Sole2019}. It is common for the physical structures of biological systems to embed intricate computational functionalities in terms of the connections, concentrations and dynamics of the diverse elements making up the system. This highlights an opportunity for the development of more efficient unconventional computers that move beyond digital logic and whose physical construction and function is tailored to the problem at hand \cite{Thomson1876, Qian2011, Greco2019, Grozinger2019, Toffoli1999, Mead1990, Chou2019, Regot2011, Benenson2012,Stepney2018}, as well as new theoretical frameworks for both digital and analog computation\cite{Philosophy2018, Computability2018, Encoding2018}. With this in mind, the challenge becomes describing how such computers might be built and how they might function to carry out useful computations.

Constructor theory is an emerging approach for thinking about and describing fundamental physics based on distinguishing between transformations of a system's state that are possible (i.e., in theory could be made to occur) and those that are not \cite{Deutsch2013}. The ability to include counterfactual statements (i.e., those transformations that are impossible) has opened up productive avenues for reconciling many previously disconnected areas of physics. In particular, there has been success in using constructor theory to integrate information theory and computation \cite{Deutsch2015}. These developments relied on information being considered as a physically instantiated entity and ensuring that the features and transformations of the physical system encoding information meets a set of requirements. These include the physical medium being able to support at least two distinct states, with the state able to be permuted and copied. Using this work as a foundation, constructor theory has also been used to show that evolution by natural selection can emerge from fundamental physics as long as the underlying system is able to physically encode digital information \cite{Marletto2015}.

Here, we take inspiration from this theory to develop what we call \emph{constructor algorithms} that specify: (i) how a physical system should be built through allowable physical transformations to create a specialized computer for a given problem; and (ii) how this constructed artefact can be used to compute solutions to a given query. By placing physical embodiment at the forefront of our algorithms, we aim to open up opportunities to consider more diverse information processing substrates and processes, such as those found in chemistry and biology \cite{Grozinger2019}. As a proof-of-concept, we consider a hypothetical robotic wire machine (RWM) that enables us to specify how a network of electrically conductive wires should be connected together in a network based on a set of input numbers and problem type. The wire network assembled encodes an unconventional computer that uses the lengths of wires, and flows and resistances of electricity through these, to efficiently solve the desired problem. We describe constructor algorithms to create unconventional computers capable of solving an NP-complete problem. Interestingly, once a particular wire network is assembled it can be used to efficiently compute solutions in $O(1)$ time, by exploiting the highly parallel nature of electrical currents. While our focus here is on demonstrating algorithms for building unconventional computers based on networks of conductive wires, we end by discussing how such constructor algorithms might translate to other physical, chemical and biological substrates to create living computers that could potentially exploit self-assembly to build such networks \cite{Greco2019, Benenson2012, Wang2017}.

\section*{Results}

\subsection*{Robotic wire machines as a basis for unconventional computers}

The basis of our unconventional computers are RWMs (\textbf{Figure 1a}). These consist of: (i) a board of parallel wires (referred to as rows) uniformly spaced at unit lengths, e.g., 1 cm; (ii) a Turing-complete programmable robot that can add additional wires connecting pairs of rows on the board in a desired pattern; and (iii) a measurement device that can introduce a fixed electrical potential across two of the rows on the board and measure any induced current. Our constructor algorithms take as input a multiset of positive integers, and from these generate a set of commands for the RWM to create a wire network able to compute solutions for a set of specified problems. To execute a computation, the user provides a query value (integer) to be tested. This value sets the length of the two connection points for the measurement device, which are then attached to relevant rows in the wire network. A fixed voltage is then applied and any electrical current measured (\textbf{Figure 1b}). If no current is observed, the wire network does not contain a path connecting those two rows. However, if a current is measured, then at least one path exists with each encoding a solution to our problem. Furthermore, we assume that the resistance of the wires connecting separated rows is significantly higher than the wires making up each row. This ensures that any electrical current that is able to pass between rows is dominated by the connecting wires and not the length of wire making up the rows themselves. Given that we known the length that each path must take (i.e., the query input) and can measure the unit length resistance of the connecting wires, from Ohm's law it is possible to calculate the number of solutions, $s$, using:
\begin{equation}
s(q, m) = \frac{V}{qR_{u}m},
\end{equation}
where $q$ is the query integer, $m$ is the measured current in amps at the voltage source, $V$ is the constant applied voltage in volts, and $R_{u}$ is the resistance of a unit length of connecting wire in ohms. It should be noted that once a wire network has been built, it can be reused as is to compute answers to many different queries --- the wire network physically embodies a set of possible computations and we use electricity to probe the solution space in parallel.

\subsection*{Encoding and efficient querying of information in wire networks}

Before describing constructor algorithms for a specific problem, it is important to understand how the resultant wire networks encode information and the function performed when querying the network. In our wire networks, positive integer numbers are encoded by the length of connecting wires. For example, a wire 5 units long encodes the number 5. Furthermore, paths of connecting wires, where the end of one wire is connected to the start of another via a shared row, encode the addition of those wire lengths. So for example, if one wire of length 2 connects rows 0 and 2, and another wire of length 3 connects rows 2 and 5, then a measurement with a query length of 5 across rows 0 to 5 will result in a flow of current and establish that addition $2 + 3 = 5$ is possible (\textbf{Figure~1b}, single path).

All of the algorithms we describe in this work build trees of connected wires that encode different combinations of additions for a set of input integers (wire lengths). The power of the wire networks comes from their ability to effectively test all of these separate additions simultaneously once the network is built, and for the number of different additions (i.e., routes through the network) to be immediately computed via the parallel nature of electronics (\textbf{Figure~1b}, multiple paths).

\subsection*{Constructor algorithm for the Subset Sum Problem}

The Subset Sum Problem (SSP) is an NP-Complete problem \cite{Karp1972} where a decision is made on whether a multiset of integers contains a subset that adds up to a target (query) integer. In this paper, we focus on a multiset containing only positive integers and a positive integer target. This version of the SSP is also NP-Complete because the problem's complexity is based on the number of subsets of the multiset, which is independent of the sign of the integers involved. We will also initially focus on basic sets without repeated elements to simplify our description. However, the same algorithm can be easily extended to multisets if redundant elements are each treated as unique in their own right. Later on we show how to optimize this algorithm for multisets if only the presence of at least one solution needs to be computed.

To demonstrate how our constructor algorithm works for SSP, consider the input set $X = \{4, 1, 5, 2\}$. The powerset $\mathbb{P}(X)$ contains all the possible subsets of $X$ that could form a solution to an SSP query and thus may require summation during a computation. Rather than carry out these sums sequentially, our constructor algorithm creates a wire network with paths embodying all possible summations for each non-empty subset. By placing the wires making up the element in each subset from end-to-end on the board, we can use the RWM's measurement device to query a target integer $q$ by connecting it to row 0 and row $q$ and measuring the induced current. As explained in the previous section, this will simultaneously test in parallel all possible sums of non-empty subsets and allow for the number of unique solutions to be ascertained. To build the wire network for an input set $X$, our constructor algorithm performs the following steps (\textbf{Figure~2a}):
\begin{enumerate}
\item For each element in $X$ generate a wire the length of its value and connect it to row 0.

\item Remove an element, $x_f$, from the remaining elements in set $X$. For every location in the current network that corresponds to element $x_f$, generate a set of wires of lengths corresponding to the remaining values in $X$, and connect them starting at these locations.

\item If $|X| > 1$ then repeat Step 2, else end as network is complete.
\end{enumerate}

\subsection*{Trade-offs in number of wires and measurements used for a computation}

For a starting set of size $n = |X|$, our standard SSP constructor algorithm will generate a network containing $n + \sum_{i = 1}^{n - 1} 2^{i-1}(n - i)=2^n-1$ wires. However, because our wire networks are physical objects, it may sometimes be useful to reduce the number of wires present to reduce costs and improve the scalability of the approach. Interestingly, this can be achieved if multiple measurements of a fixed wire network are possible. This then enables a trade-off between the number of measurements and size of the wire network produced.

To understand why this is the case, it is helpful to recognise the self-similarity of sub-networks that are present in the wire networks built using the standard SSP algorithm (see shaded regions in \textbf{Figure~2b}). It is clear that two large and identical sub-networks exist, which calculate the SSP for the set excluding the first element 5. Therefore, if two measurements are possible for the same query, the entire sub-network that tests the set $\{2, 1, 3, 4\}$ that starts at row 0 can be removed and an additional measurement made starting at the row where the copy of the sub-network begins (row 5 in this example). This allows a network containing only $1+[(n-1) + \sum_{i = 1}^{(n - 1)-1} 2^{i-1}((n-1) - i)]=1+2^{n-1}-1=2^{n-1}$ wires (the main sub-network shown in the darkest grey in \textbf{Figure 2b}, plus a wire for the first element in the set) to be used to compute the SSP if two measurements are possible (\textbf{Figure 2c}).

A further benefit of the wire networks is their ability to be easily reused. Performing multiple queries for the same starting set merely requires a change to the query length used for the measurement of electrical currents. Therefore, while sequentially building the network is relatively slow, querying for both one measurement and two measurement situations is always fast: $O(1)$.

\subsection*{Growing wire networks to accommodate new input set values}

Once a wire network is built, if new elements are added to the input they could be accommodated by rebuilding the network from scratch. This would impart a large time cost and much of the new network would be identical to the original one. An alternative approach is to allow for the wire networks to ``grow'' via defined transformations. In this case, the addition of a new element in the input set can be easily integrated into an existing network by taking all occurrences of the last element added to the tree and adding two new wires to the start row and end row of each last element occurrence with lengths that equal to the new element's value (\textbf{Figure~3}). This transformation effectively adds branches to all possible sub-solutions that include the new element.

\subsection*{Optimizing wire networks for multisets}

So far, we have treated the repeated elements of a multiset as though they were unique elements in their own right. This ensures the networks created contain all possible summations as unique routes through the wire network. However, if we are only interested in whether a solution exists and not concerned with the number of unique solutions, then simpler wire networks exist to carry out this computation. This stems from the fact that repeated elements lead to redundancy in the wire network where paths incorporate the elements with identical values at different steps of the paths created (because each element is treated as unique by our original algorithm). This is evident by looking at the structure of a wire network built for an input with an element value that is repeated (\textbf{Figure~4a}). At a particular level (i.e., row) of the network, there is repetition across sub-networks, allowing some to be removed without affecting the output of the computation.

To generate an optimised network it is possible to retrospectively analyse the network and remove identical sub-networks at each level. However, a simpler approach is to not produce these sub-networks when the wire network is first being produced. To do this we can adapt our previous algorithm to filter out these unnecessary elements as they arise. The constructor algorithm for a wire network where only the presence of at least one solution is required and two measurements are possible is given by:
\begin{enumerate}
\item Remove an element, $x_f$, from $X$, create a wire the length of its value and connect it to row 0.

\item For the remaining elements in $X$, generate a single wire for each value (i.e., two or more elements with the same value only leads to a single wire) and connect them to the row at the end of the wire added in Step 1.

\item Remove another element, $x_f$, from the remaining elements in set $X$. For every wire in the current network that corresponds to the element $x_f$, generate a single wire for each value (i.e., two or more elements with the same value only leads to a single wire) and connect them all starting at these locations.

\item If $|X| > 1$ then repeat Step 3, else end as network is complete.
\end{enumerate}
An example of the steps involved in this algorithm are shown in \textbf{Figure~4b}.

\section*{Discussion}

We typically consider information and computation as abstract concepts. However, the success of constructor theory to unify these ideas in theoretical physics demonstrates that real-world constraints need not hamper our ability to express these ideas \cite{Deutsch2015, Marletto2015, Deutsch2013}. In this work, we have introduced the concept of constructor algorithms as a means to describe the physical transformations that take place when building an unconventional computing device able to solve the NP-complete Subset Sum Problem. Our focus has been to demonstrate how this approach offers a framework for thinking about unconventional computing by constraining our computers to being built as simple wire networks via a programmable RWM. Our constructor algorithms are able to tailor the RWM’s actions (i.e., physical transformations of the wire network) and thus the structure of the resultant network for a specific problem at hand. Once built, this network can then be used to immediately compute the answer to any number of queries by exploiting the lengths of paths through this network and the parallel nature of electrical currents.

We chose simple wire networks as our computing medium to highlight that an alternative physical encoding of information (e.g., numbers encoded as the lengths of wires) can lead to new ways that a computer can be physically built and parallel computations performed. In this case, the wire networks encode the full solution space of the problem and are able to exploit changes in electrical current to provide the number of unique solutions at no extra cost.

Being physical entities, as our wire networks grow in complexity, we are likely to hit practical limits for their use. We already assume that the wires for each row have significantly less electrical resistance than those wires connecting rows, which ensures the major determinant of the electrical resistance comes from the connecting wires. This avoids issues where the physical placement of wires on a single row generates large horizontal distances that would affect our ability to determine if multiple paths existed. However, there will be physical limits as the size of the input multiset grows causing this assumption to break. It should be noted, that even if this assumption does not hold, we are still able to test whether at least one solution exists by monitoring any flow of a current to solve the classical SSP.

In addition to wire placement, the physical size of the networks built will also cause problems as their complexity grows. To compensate for this, we can scale down the unit length and thus the amount of wire needed, as well as the diameter of each wire. Such changes would also require enhancement to the RWM to allow for the precise handling of tiny components. Richard Feynman is famous for stating that ``there’s plenty of room at the bottom'' \cite{Feynman1959}, but controlling what happens at the micro- or nano-scales is hugely challenging. One potential way of achieving such a goal is to make use of molecular self-assembly (e.g., DNA origami \cite{Wang2017}), allowing a wire network to effectively build itself. This parallel assembly would not only scale-down the size of our networks, but also significantly improve the assembly time and reduce the overall complexity of creating and executing computations using this approach.

In summary, all life senses, responds, and adapts to its environment. Thinking about whether algorithms might exist for describing how organisms develop to implement personalised computational devices throughout their lives may be an important step towards new approaches to unconventional computers that evolve with use. We believe that constructor algorithms provide a framework to support this goal, placing the physical embodiment of computation centre-stage and enabling input data to affect how a computer’s internal structure is built, and thus, the flow of information when computations occur. This goes beyond the sequential state-based architectures of general-purpose electronic computers we all use today, offering inspiration for new ways of harnessing diverse physical, chemical and biological processes for our growing computational needs.

\section*{Acknowledgements}

L.S. was supported by U.S.A. National Science Foundation Grant No. 2117377. T.E.G. was supported by a Royal Society University Research Fellowship grant UF160357, and a Turing Fellowship from The Alan Turing Institute under the EPSRC grant EP/N510129/1. Any opinions, findings, and conclusions or recommendations expressed in this publication are those of the authors and do not necessarily reflect the views of the National Science Foundation or other funders. The funders also had no role in study design, data collection and analysis, and decision to publish or preparation of the manuscript.

\section*{Author Contributions}

T.M. conceived the project, developed the constructor algorithms, and proved the mathematical results. L.S. and T.E.G. provided guidance and support during development of the research. All authors contributed to the writing and editing of the manuscript.

\section*{Conflicts of interest}

None declared.

\newpage
\section*{Figures and captions}

\begin{center}
\includegraphics[width=12.4cm]{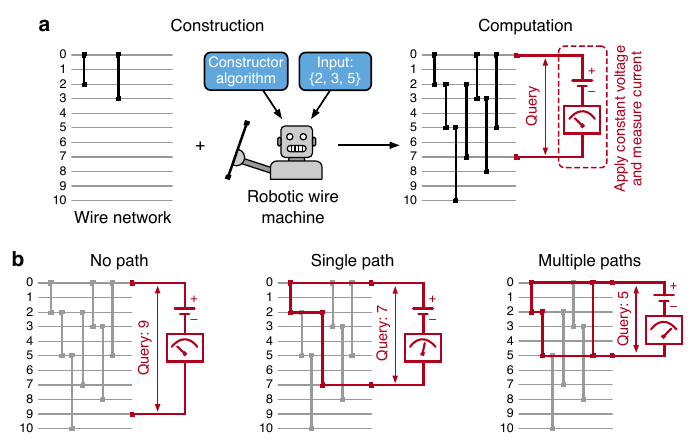}
\end{center}
\noindent\textsf{\textbf{Figure 1: Overview of a robotic wire machine and constructor algorithm.}}
(\textbf{\textsf{a}}) Our constructor algorithms take as input a multiset of non-negative integers and then control how a robotic wire machine (RWM) creates a wire network (our non-conventional computer) tailored for queries/computations related to that input. The RWM is able to connect pairs of parallel, uniformly spaced horizontal wires (rows) with insulated wires of defined lengths running vertically to encode the solution space. A computation is performed using this generated circuit and connecting a measurement device (containing an electrical source of constant voltage and element to measure the current) with a length equal to the user's query. If an electrical current is detected, then a solution exists.
(\textbf{\textsf{b}}) Examples of the three possible outputs of a computation: no solution (path), a single solution, or multiple solutions. By using Ohm's law and the fact that our measurement applies a fixed voltage and that we know the resistance of a unit length of connecting wire, changes in current can be used to calculate the number of solutions present. Possible paths are shown in red.

\newpage
\begin{center}
\includegraphics[width=\textwidth]{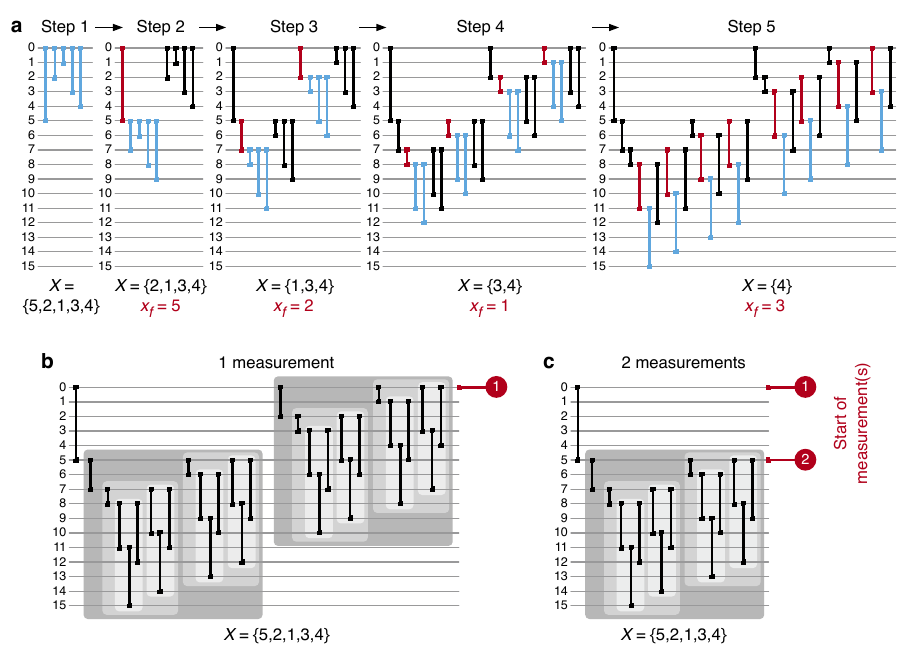}
\end{center}
\noindent\textsf{\textbf{Figure 2: Constructor algorithm for the Subset Sum Problem.}}
(\textbf{\textsf{a}}) Steps involved in creating the wire network for the input set $X = \{5, 2, 1, 3, 4\}$. At each step, the element removed from the set is shown below the wire network in red, as well as the remaining elements in the input set. Blue wires denote new wires added at that step, red wires denote those wires to which new links are being made.
(\textbf{\textsf{b}}) The final wire network can be queried using a single measurement starting at row 0.
(\textbf{\textsf{c}}) Because of the self similar nature of the wire network (similar sub-networks highlighted in shades of grey), if two measurements are possible then the wire network can be simplified to include only one of the largest repeated sub-networks (shown in dark grey). In this case though a measurement needs to be taken at row 0 and at the row where the top of the repeated sub-network is placed.

\newpage
\begin{center}
\includegraphics[width=14cm]{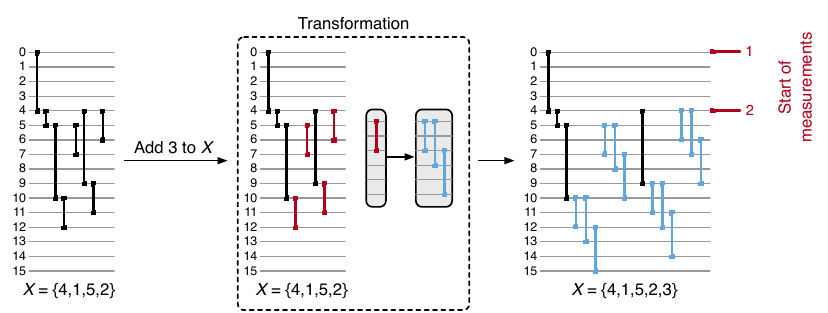}
\end{center}
\noindent\textsf{\textbf{Figure 3: Adapting a wire networks to new input elements.}}
In this example an initial wire network requiring two measurements for the input set $X = \{4, 1, 5, 2\}$ is updated to include a new element, 3. A transformation is used to replace all of the last inserted elements (highlighted in red) with sets of wires that have wires corresponding to the new element placed at the start row and end row of the last inserted elements.

\newpage
\begin{center}
\includegraphics[width=16.5cm]{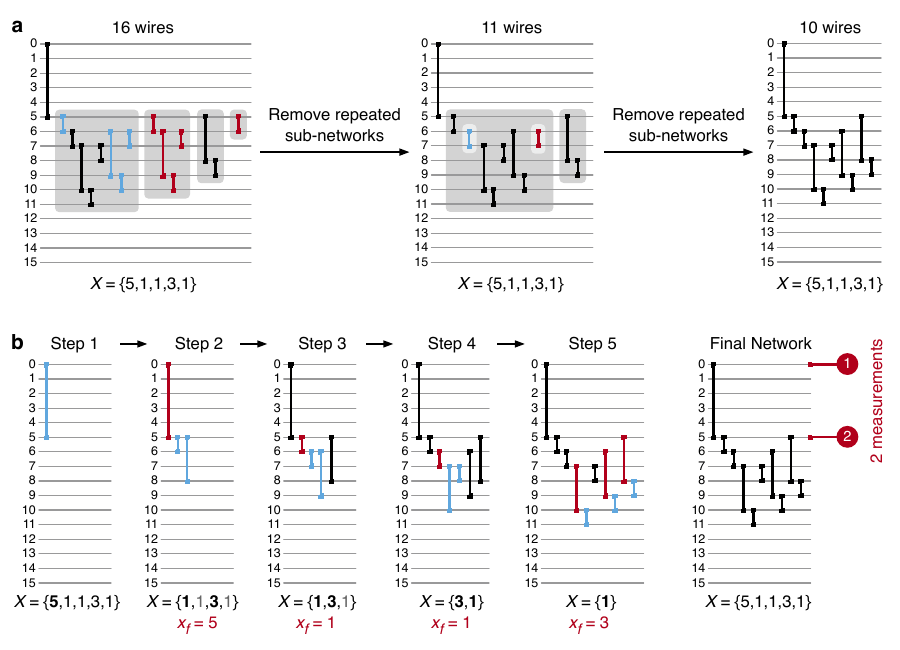}
\end{center}
\noindent\textsf{\textbf{Figure 4: Optimising wire networks for multisets and repeated values if only the presence of at least one solution required.}}
(\textbf{\textsf{a}}) A wire network for the input multiset $X = \{5, 1, 1, 3, 1\}$ containing repeated elements is shown using our original constructor algorithm. Sub-networks generated for each element in the input set are highlighted in grey. Repeated sub-networks (red) that occur within the largest sub-network (blue) can be removed without affecting the correctness of the output if only the presence of at least one solution is required. This optimisation can be continued to deeper levels within the network until no further levels exist.
(\textbf{\textsf{b}}) Optimised constructor algorithm for multisets where only the presence of at least one solution is required. Wires being added to the network are highlighted in blue, current selected wires are highlighted in red, and elements in bold in the set $X$ below each wire network correspond to those for which wires are created (repeated values only have a single wire created). Computations of these networks require 2 measurements.

\clearpage
\bibliography{references.bib}

\begin{thebibliography}{10}
\expandafter\ifx\csname url\endcsname\relax
  \def\url#1{\texttt{#1}}\fi
\expandafter\ifx\csname urlprefix\endcsname\relax\def\urlprefix{URL }\fi
\providecommand{\bibinfo}[2]{#2}
\providecommand{\eprint}[2][]{\url{#2}}

\bibitem{Hameed2010}
\bibinfo{author}{Hameed, R.} \emph{et~al.}
\newblock \bibinfo{title}{Understanding sources of inefficiency in
  general-purpose chips}.
\newblock In \emph{\bibinfo{booktitle}{Proceedings of the 37th Annual
  International Symposium on Computer Architecture}}, ISCA '10,
  \bibinfo{pages}{37–47} (\bibinfo{publisher}{Association for Computing
  Machinery}, \bibinfo{year}{2010}).

\bibitem{Bourret2002}
\bibinfo{author}{Bourret, R.~B.} \& \bibinfo{author}{Stock, A.~M.}
\newblock \bibinfo{title}{Molecular information processing: lessons from
  bacterial chemotaxis}.
\newblock \emph{\bibinfo{journal}{Journal of Biological Chemistry}}
  \textbf{\bibinfo{volume}{277}}, \bibinfo{pages}{9625--9628}
  (\bibinfo{year}{2002}).

\bibitem{Brody2023}
\bibinfo{author}{Brody, D.~C.} \& \bibinfo{author}{Trewavas, A.~J.}
\newblock \bibinfo{title}{Biological efficiency in processing information in
  green plants}.
\newblock \emph{\bibinfo{journal}{Proceedings of the Royal Society A:
  Mathematical, Physical and Engineering Sciences}}
  \textbf{\bibinfo{volume}{479}}, \bibinfo{pages}{20220809}
  (\bibinfo{year}{2023}).

\bibitem{Berridge2001}
\bibinfo{author}{Berridge, M.~J.}
\newblock \emph{\bibinfo{title}{The Versatility and Complexity of Calcium
  Signalling}}, \bibinfo{pages}{52--67} (\bibinfo{publisher}{John Wiley \&
  Sons, Ltd}, \bibinfo{year}{2001}).

\bibitem{Adamatzky2018}
\bibinfo{author}{Adamatzky, A.}
\newblock \bibinfo{title}{Towards fungal computer}.
\newblock \emph{\bibinfo{journal}{Interface Focus}}
  \textbf{\bibinfo{volume}{8}}, \bibinfo{pages}{20180029}
  (\bibinfo{year}{2018}).

\bibitem{Sole2019}
\bibinfo{author}{Solé, R.}, \bibinfo{author}{Moses, M.} \&
  \bibinfo{author}{Forrest, S.}
\newblock \bibinfo{title}{Liquid brains, solid brains}.
\newblock \emph{\bibinfo{journal}{Philosophical Transactions of the Royal
  Society B: Biological Sciences}} \textbf{\bibinfo{volume}{374}},
  \bibinfo{pages}{20190040} (\bibinfo{year}{2019}).

\bibitem{Thomson1876}
\bibinfo{author}{Thomson, W.}
\newblock \bibinfo{title}{Iv. on an instrument for calculating ($\int\theta (x)
  \phi (x) dx$), the integral of the product of two given functions}.
\newblock \emph{\bibinfo{journal}{Proceedings of the Royal Society of London}}
  \textbf{\bibinfo{volume}{24}}, \bibinfo{pages}{266--268}
  (\bibinfo{year}{1876}).

\bibitem{Qian2011}
\bibinfo{author}{Qian, L.}, \bibinfo{author}{Winfree, E.} \&
  \bibinfo{author}{Bruck, J.}
\newblock \bibinfo{title}{Neural network computation with dna strand
  displacement cascades}.
\newblock \emph{\bibinfo{journal}{Nature}} \textbf{\bibinfo{volume}{475}},
  \bibinfo{pages}{368--372} (\bibinfo{year}{2011}).

\bibitem{Greco2019}
\bibinfo{author}{Greco, F.~V.}, \bibinfo{author}{Tarnowski, M.~J.} \&
  \bibinfo{author}{Gorochowski, T.~E.}
\newblock \bibinfo{title}{Living computers powered by biochemistry}.
\newblock \emph{\bibinfo{journal}{The Biochemist}}
  \textbf{\bibinfo{volume}{41}}, \bibinfo{pages}{14--18}
  (\bibinfo{year}{2019}).

\bibitem{Grozinger2019}
\bibinfo{author}{Grozinger, L.} \emph{et~al.}
\newblock \bibinfo{title}{Pathways to cellular supremacy in biocomputing}.
\newblock \emph{\bibinfo{journal}{Nature Communications}}
  \textbf{\bibinfo{volume}{10}}, \bibinfo{pages}{5250} (\bibinfo{year}{2019}).

\bibitem{Toffoli1999}
\bibinfo{author}{Toffoli, T.}
\newblock \bibinfo{title}{Programmable matter methods}.
\newblock \emph{\bibinfo{journal}{Future Generation Computer Systems}}
  \textbf{\bibinfo{volume}{16}}, \bibinfo{pages}{187--201}
  (\bibinfo{year}{1999}).

\bibitem{Mead1990}
\bibinfo{author}{Mead, C.}
\newblock \bibinfo{title}{Neuromorphic electronic systems}.
\newblock \emph{\bibinfo{journal}{Proceedings of the IEEE}}
  \textbf{\bibinfo{volume}{78}}, \bibinfo{pages}{1629--1636}
  (\bibinfo{year}{1990}).

\bibitem{Chou2019}
\bibinfo{author}{Chou, J.}, \bibinfo{author}{Bramhavar, S.},
  \bibinfo{author}{Ghosh, S.} \& \bibinfo{author}{Herzog, W.}
\newblock \bibinfo{title}{Analog coupled oscillator based weighted ising
  machine}.
\newblock \emph{\bibinfo{journal}{Scientific reports}}
  \textbf{\bibinfo{volume}{9}}, \bibinfo{pages}{14786} (\bibinfo{year}{2019}).

\bibitem{Regot2011}
\bibinfo{author}{Regot, S.} \emph{et~al.}
\newblock \bibinfo{title}{Distributed biological computation with multicellular
  engineered networks}.
\newblock \emph{\bibinfo{journal}{Nature}} \textbf{\bibinfo{volume}{469}},
  \bibinfo{pages}{207--211} (\bibinfo{year}{2011}).

\bibitem{Benenson2012}
\bibinfo{author}{Benenson, Y.}
\newblock \bibinfo{title}{Biomolecular computing systems: principles, progress
  and potential}.
\newblock \emph{\bibinfo{journal}{Nature Reviews Genetics}}
  \textbf{\bibinfo{volume}{13}}, \bibinfo{pages}{455--468}
  (\bibinfo{year}{2012}).

\bibitem{Stepney2018}
\bibinfo{editor}{Stepney, S.}, \bibinfo{editor}{Rasmussen, S.} \&
  \bibinfo{editor}{Amos, M.} (eds.).
\newblock \emph{\bibinfo{title}{Computational Matter}}
  (\bibinfo{publisher}{Springer}, \bibinfo{year}{2018}).

\bibitem{Philosophy2018}
\bibinfo{author}{Konkoli, Z.} \emph{et~al.}
\newblock \emph{\bibinfo{title}{Computational Matter}}, chap.
  \bibinfo{chapter}{Philosophy of Computation}, \bibinfo{pages}{153--184}
  (\bibinfo{publisher}{Springer}, \bibinfo{year}{2018}).

\bibitem{Computability2018}
\bibinfo{author}{Broersma, H.}, \bibinfo{author}{Stepney, S.} \&
  \bibinfo{author}{Wendin, G.}
\newblock \emph{\bibinfo{title}{Computational Matter}}, chap.
  \bibinfo{chapter}{Computability and Complexity of Unconventional Computing
  Devices}, \bibinfo{pages}{185--229} (\bibinfo{publisher}{Springer},
  \bibinfo{year}{2018}).

\bibitem{Encoding2018}
\bibinfo{author}{McCaskill, J.~S.}, \bibinfo{author}{Miller, J.~F.},
  \bibinfo{author}{Stepney, S.} \& \bibinfo{author}{Wills, P.~R.}
\newblock \emph{\bibinfo{title}{Computational Matter}}, chap.
  \bibinfo{chapter}{Encoding and Representation of Information Processing in
  Irregular Computational Matter}, \bibinfo{pages}{231--248}
  (\bibinfo{publisher}{Springer}, \bibinfo{year}{2018}).

\bibitem{Deutsch2013}
\bibinfo{author}{Deutsch, D.}
\newblock \bibinfo{title}{Constructor theory}.
\newblock \emph{\bibinfo{journal}{Synthese}} \textbf{\bibinfo{volume}{190}},
  \bibinfo{pages}{4331--4359} (\bibinfo{year}{2013}).

\bibitem{Deutsch2015}
\bibinfo{author}{Deutsch, D.} \& \bibinfo{author}{Marletto, C.}
\newblock \bibinfo{title}{Constructor theory of information}.
\newblock \emph{\bibinfo{journal}{Proceedings of the Royal Society A:
  Mathematical, Physical and Engineering Sciences}}
  \textbf{\bibinfo{volume}{471}}, \bibinfo{pages}{20140540}
  (\bibinfo{year}{2015}).

\bibitem{Marletto2015}
\bibinfo{author}{Marletto, C.}
\newblock \bibinfo{title}{Constructor theory of life}.
\newblock \emph{\bibinfo{journal}{Journal of The Royal Society Interface}}
  \textbf{\bibinfo{volume}{12}}, \bibinfo{pages}{20141226}
  (\bibinfo{year}{2015}).

\bibitem{Wang2017}
\bibinfo{author}{Wang, P.}, \bibinfo{author}{Meyer, T.~A.},
  \bibinfo{author}{Pan, V.}, \bibinfo{author}{Dutta, P.~K.} \&
  \bibinfo{author}{Ke, Y.}
\newblock \bibinfo{title}{The beauty and utility of dna origami}.
\newblock \emph{\bibinfo{journal}{Chem}} \textbf{\bibinfo{volume}{2}},
  \bibinfo{pages}{359--382} (\bibinfo{year}{2017}).

\bibitem{Karp1972}
\bibinfo{author}{Karp, R.~M.}
\newblock \emph{\bibinfo{title}{Reducibility among Combinatorial Problems}},
  \bibinfo{pages}{85--103} (\bibinfo{publisher}{Springer US},
  \bibinfo{year}{1972}).

\bibitem{Feynman1959}
\bibinfo{author}{Feynman, R.}
\newblock \bibinfo{title}{Plenty of room at the bottom}.
\newblock \emph{\bibinfo{journal}{Transcript of a talk given to the American
  Physical Society}}  (\bibinfo{year}{1959}).

\end{thebibliography}
\end{document}